\documentclass[aps,prl,onecolumn,11pt,superscriptaddress,floatfix]{revtex4-2}

\usepackage{graphicx}
\usepackage{amsmath}
\usepackage{amssymb}
\usepackage{bm}

\begin{document}

\title{Identification and Structural Decomposition of Hidden Defect Configurations: A Case Study of Charged Oxygen Divacancies in HfO$_2$}

\author{Hyunjin Lee}
\affiliation{Department of Energy Science and Engineering, Daegu Gyeongbuk Institute of Science \& Technology (DGIST), Daegu, Republic of Korea}
\author{Yoon Kyeung Lee}
\affiliation{Division of Advanced Materials Engineering, Jeonbuk National University, Jeonju 54896, Republic of Korea}
\author{Keun Heo}
\email{kheo@jbnu.ac.kr}
\affiliation{School of Semiconductor and Chemical Engineering, Jeonbuk National University, Jeonju 54896, Korea}
\author{Shi Liu}
\email{liushi@westlake.edu.cn}
\affiliation{Department of Physics, School of Science, Westlake University, Hangzhou, China}
\author{Taehun Lee}
\email{taehun.lee@dgist.ac.kr}
\affiliation{Department of Energy Science and Engineering, Daegu Gyeongbuk Institute of Science \& Technology (DGIST), Daegu, Republic of Korea}
\date{\today}

\begin{abstract}
Extensive configurational searches have revealed hidden defect structures in crystalline compounds, yet how their local bonding arrangements modify the surrounding atomic environment remains less well understood. We examine this question in hafnium oxide (HfO$_2$), which hosts energetically competing monoclinic (M), tetragonal (T), and polar orthorhombic (PO) polymorphs. For the $c$-axis oxygen-divacancy complex ($T_c$), a local-distortion search with density-functional theory (DFT) relaxation under tetragonal-lattice confinement identifies a hidden family of low-energy configurations ($T_c^{*}$). For the fully charged complex ($2\mathrm{V}_{\mathrm{O}}^{2+}$), structural decomposition using Smooth Overlap of Atomic Positions (SOAP) descriptors reveals mixed M-, T-, and PO-like character, supported by independent analyses of Hf coordination and bond geometry. The idealized $T_c$ reference connects barrierlessly to the lowest-energy mixed-motif $T_c^{*}$ configuration, lowering the energy by $18.4$~meV/f.u.\ within the confined landscape. After releasing the lattice constraint, we compare transformation pathways to M and PO using solid-state nudged elastic band calculations with cell-relaxed endpoints. The respective barriers increase from $6.9$ and $10.1$~meV/f.u.\ for the idealized reference to $26.9$ and $15.3$~meV/f.u.\ for the $T_c^{*}$-derived structure. The larger increase for M makes PO, rather than M, the lower-barrier branch. This case study shows how structural decomposition connects hidden defect configurations with changes in the surrounding atomic environment and kinetic branching between competing polymorphs.
\end{abstract}

\maketitle
\clearpage

Defects in crystalline compounds can induce hidden bonding arrangements that are difficult to anticipate from the ideal crystal structure. Considerable effort has focused on identifying such arrangements through configurational searches~\cite{Mosquera2023,Mosquera2022,Arrigoni2021}, but a reconstructed defect does not stop at its own coordination shell, and how far its influence extends into the surrounding lattice is rarely resolved. In polymorphic materials, local environments reconstructed around defects can retain structural features of several competing phases, making their polymorphic character difficult to resolve. Addressing this gap requires both identifying reconstructed configurations and characterizing their local coordination and structural similarity to competing polymorphic reference motifs. We examine this challenge in hafnium oxide (HfO$_2$), using structural decomposition to connect hidden defect configurations with local structural changes and kinetic branching between competing polymorphs.

Ferroelectricity in HfO$_2$ is associated with the non-centrosymmetric polar orthorhombic phase (space group $Pca2_1$, hereafter \textbf{PO})~\cite{Sang2015}. Under ambient bulk conditions, HfO$_2$ adopts the monoclinic $P2_1/c$ phase (\textbf{M}), whereas the tetragonal $P4_2/nmc$ phase (\textbf{T}) is a high-temperature polymorph~\cite{Wang1992}. Distinguishing their local structural character is challenging because M and PO share sevenfold Hf coordination and comparable Hf--O distances while differing in the symmetry of their bonding networks~\cite{Yuan2023}. In thin films, surface and grain-boundary energies, dopants, and epitaxial or electrode-induced confinement modify the relative stability and transformation kinetics of these polymorphs, making T accessible beyond its bulk stability range~\cite{Materlik2015,Batra2016,Park2017,Park2017Nanoscale,Kunneth2017}. T also provides a common structural precursor to M and PO~\cite{Huan2014}: first-principles studies identify low-barrier pathways connecting T to both phases~\cite{Liu2019}, with the preferred route depending on confinement~\cite{Dogan2019}. Although M is the bulk ground state and PO is metastable under strain-free conditions, the relative kinetics of these T-derived pathways can influence phase selection. This raises the question of how point defects and their charge states reconstruct the local environment and alter the competing M and PO pathways.

Oxygen vacancies (V$_{\mathrm{O}}$) are particularly relevant because they are key intrinsic defects in HfO$_2$-based films~\cite{Chouprik2021,Guha2007}. Changes in local Hf coordination and polymorph stability provide a microscopic route through which V$_{\mathrm{O}}$ affects the M--PO competition~\cite{Zhou2019,He2021}. Experimental and computational studies have linked vacancy redistribution and ordering to structural evolution, ferroelectric enhancement, and altered polymorph stability~\cite{Nukala2021,Grimley2016,Kang2022,Mittmann2020,Rushchanskii2021,Ma2023}. For the positively charged, $c$-axis-aligned V$_{\mathrm{O}}$--O--V$_{\mathrm{O}}$ configuration denoted $T_c$, reported $T_c{\to}$PO and $T_c{\to}$M barriers of $24$ and $54$~meV/f.u., respectively, indicate a polar kinetic preference~\cite{Ma2023}. These pathways were referenced to an idealized, bulk-like $T_c$ geometry. Because defect configurations can undergo substantial local coordination rearrangements~\cite{Mosquera2023,Mosquera2022}, assessing this preference requires establishing the reconstructed configurations accessible within the charged $T_c$ topology and determining how their structural character affects kinetic branching.

Here, we combine local-distortion sampling, density-functional theory based geometric relaxation, and SOAP-based structural decomposition to resolve this configurational landscape. We generate $980$ locally distorted starting structures across total charge states $q=0,+1,+2,+3,$ and $+4$ and relax them under tetragonal-lattice confinement, separating local configurational rearrangements from geometric changes. Hf-coordination analysis characterizes changes in the atomic environment around the divacancy, while SOAP fingerprints~\cite{Bartok2013} quantify structural similarity through projection onto an M/T/PO reference simplex. The resulting weights describe similarity to the reference motifs rather than bulk phase fractions. We then release the lattice constraint for selected configurations and evaluate the competing polymorphic pathways using solid-state nudged elastic band (SS-NEB) calculations with cell-relaxed endpoints~\cite{Sheppard2012,Henkelman2000}.

The search reveals a family of hidden, lower-energy configurations, denoted $T_c^{*}$, that preserve the $c$-axis vacancy alignment and retain mixed M-, T-, and PO-like structural character. For the fully charged complex ($q=+4$), the idealized reference connects barrierlessly to the lowest-energy mixed-motif $T_c^{*}$ configuration, energetically gaining $18.4$~meV/f.u.\ under lattice confinement. Its stabilization and mixed structural character persist after variable-cell relaxation. Within the present SS-NEB workflow, an additional M channel involving oxygen displacements distributed across the supercell lowers the idealized-reference M barrier to $6.9$~meV/f.u., below the PO barrier of $10.1$~meV/f.u. Using the corresponding cell-relaxed $T_c^{*}$-derived structure changes the M and PO barriers to $26.9$ and $15.3$~meV/f.u., respectively. The larger increase in the M barrier makes PO, rather than M, the lower-barrier branch. Structural decomposition thus connects the hidden configurational landscape to changes in local coordination and to the kinetic consequences of defect-induced reconstruction.

The $c$-axis di-V$_{\mathrm{O}}$ motif ($T_c$) was constructed by
sequentially removing two $c$-axis-aligned oxygen atoms from a
$2{\times}2{\times}2$ $P4_2/nmc$ HfO$_2$ supercell (Hf$_{32}$O$_{64}$, $96$ atoms;
long axis $a\,{=}\,10.45$~\AA{}, short axes $b\,{=}\,c\,{=}\,10.15$~\AA{}) while
retaining the intervening oxygen atom, leaving Hf$_{32}$O$_{62}$ and
a V$_{\mathrm{O}}$ concentration of
$3.125\%$.
Total charge states $q=0$ to
$+4$ were considered. Positive charge states were modeled by removing
$q$ electrons with a uniform compensating background charge~\cite{He2021}.
At each total charge state, local bond distortions were applied independently around
each V$_{\mathrm{O}}$ center by scaling the initial distance between each V$_{\mathrm{O}}$ center and its
neighboring atoms by a factor $1+\delta$, with $\delta$ ranging from $-60\%$ to $+60\%$ in
$10\%$ increments~\cite{Mosquera2022,Kavanagh2024}.
Throughout this local search, the lattice vectors were held fixed to those of the tetragonal cell. The consequences of releasing the lattice constraint are examined separately below. The full distortion protocol is described in the Supplemental Material~\cite{SM}, Sec.~I.

The distorted configurations generated were then relaxed with density-functional
theory (DFT) where DFT calculations were performed using the Vienna \textit{ab initio}
simulation package (VASP) with projector augmented-wave potentials and the
Perdew--Burke--Ernzerhof functional~\cite{Kresse1996,Kresse1999,Perdew1996}. Calculations
were spin polarized and used a $600$~eV plane-wave cutoff and a $\Gamma$-centered $k$ mesh
with $0.30$~\AA$^{-1}$ spacing. Structures were relaxed with a force tolerance of
$0.01$~eV/\AA{}. The relaxed configurations were characterized using Smooth Overlap
of Atomic Positions (SOAP) descriptors~\cite{Bartok2013}. Their fingerprints were used
both to construct a two-dimensional principal-component landscape and to quantify local M-, T-, and
PO-like motif similarity through projection onto an M/T/PO reference simplex.
Transition kinetic pathways from the selected $q=+4$ endpoints were then obtained with the
solid-state nudged elastic band (SS-NEB) method~\cite{Sheppard2012}, which
accounts for both atomic displacements and cell degrees of freedom along the path, building on the climbing-image nudged elastic band method~\cite{Henkelman2000}. Full SOAP parameters,
projection details, and transition-path settings are given in the Supplemental Material~\cite{SM}, Secs.~II, III, and~VII.

\begin{figure*}[ht!]
\centering
\includegraphics[width=0.95\textwidth]{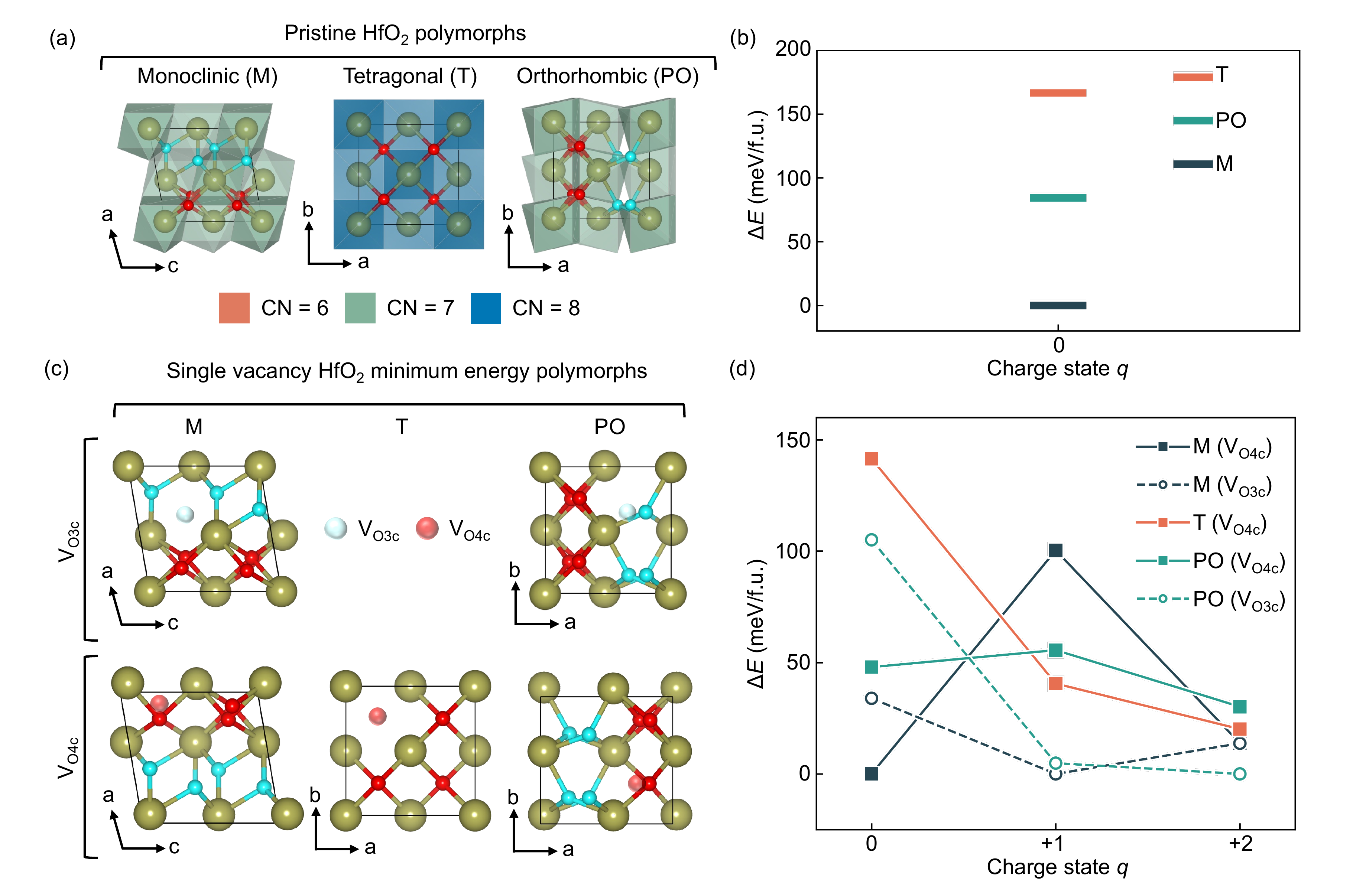}
\caption{(a) Pristine HfO$_2$ polymorphs, monoclinic (\textbf{M}), tetragonal (\textbf{T}), and polar
orthorhombic (\textbf{PO}), colored by Hf coordination number (CN${=}6,7,8$); oxygen atoms are
colored by site type (O$_{\mathrm{3c}}$, cyan; O$_{\mathrm{4c}}$, red). (b) Relative energy
$\Delta E$ (meV/f.u.) of the pristine polymorphs in the neutral charge state, referenced to
the M ground state.
(c) Minimum-energy structures of the three-coordinated (V$_{\mathrm{O3c}}$) and
four-coordinated (V$_{\mathrm{O4c}}$) oxygen vacancies in each polymorph. (d) Relative
energy of the vacancy-containing polymorphs as a function of charge state $q=0,+1,+2$ for both vacancy sites, referenced within each charge state to its lowest-energy structure.}
\label{fig:fig1}
\end{figure*}

Pristine bulk HfO$_2$ polymorphs and single V$_{\mathrm{O}}$ structures were analyzed first to establish
the structural and energetic baseline for the di-V$_{\mathrm{O}}$ search.
Figure~\ref{fig:fig1}(a) shows the three relevant HfO$_2$ polymorphs (M, T, PO) together
with their characteristic Hf coordination polyhedra, color-coded by coordination
number (CN${=}6,7,8$). In the pristine
cells, M and PO are sevenfold coordinated around Hf, with mean Hf--O
distances near $2.15$~\AA{}, whereas T is uniformly eightfold coordinated with longer and
more broadly distributed Hf--O bonds near $2.23$~\AA{}.
Each polymorph hosts
oxygen sites classified as three-coordinated (O$_{\mathrm{3c}}$) or four-coordinated
(O$_{\mathrm{4c}}$). M and PO contain both site types, whereas T contains only
O$_{\mathrm{4c}}$ sites. The computed relative
energies confirm the M ground state, with PO and T
lying $84$ and $167$~meV/f.u.\ above it (Fig.~\ref{fig:fig1}(b)), consistent with the
energetic ordering in prior first-principles
studies~\cite{Huan2014,Materlik2015}.
The corresponding Hf--O
bond-length, O--Hf--O angle, and Hf-coordination statistics are summarized in the
Supplemental Material~\cite{SM}, Sec.~IV.

Although locally distorted configurations were sampled for each symmetry-distinct V$_{\mathrm{O}}$ site across the M, T, and PO polymorphs, all distorted inputs relaxed back to their respective minimum-energy structures shown in Fig.~\ref{fig:fig1}(c). Specifically, the search included V$_{\mathrm{O3c}}$ and V$_{\mathrm{O4c}}$ in M and PO and V$_{\mathrm{O4c}}$ in T, but identified no additional low-energy minima.
The apparent PO stabilization in the $1{\times}1{\times}1$ single V$_{\mathrm{O}}$ supercell
(a high $12.5\%$ V$_{\mathrm{O}}$ concentration) occurs only
for $q=+2$ (Fig.~\ref{fig:fig1}(d)), consistent with prior calculations showing that
V$_{\mathrm{O}}^{2+}$ can stabilize PO over M at sufficiently high vacancy
concentrations~\cite{He2021}.
Quantitatively, increasing the vacancy charge state from $q=0$ to $+2$ progressively narrows the energy spread among the vacancy-containing polymorphs in this single-V$_{\mathrm{O}}$ supercell, from $142$~meV/f.u.\ at $q=0$ to $100$~meV/f.u.\ at $q=+1$ and $30$~meV/f.u.\ at $q=+2$, where PO becomes lower in energy than M.

Having established the single V$_{\mathrm{O}}$ baseline, we next extend the configurational search to the $c$-axis di-V$_{\mathrm{O}}$ complex ($T_c$) in a $2{\times}2{\times}2$ supercell (Supplemental Material~\cite{SM}, Sec.~I).
Unlike the isolated V$_{\mathrm{O}}$, under tetragonal-lattice confinement, the local-distortion search, performed while preserving the $c$-axis V$_{\mathrm{O}}$--O--V$_{\mathrm{O}}$ alignment, reveals a hidden family of lower-energy configurations, denoted $T_c^{*}$, at every charge state from $q=0$ to $+4$.
To enable direct comparison with the charged-divacancy kinetics reported in Ref.~\cite{Ma2023}, we focus below on the fully charged di-V$_{\mathrm{O}}$ complex, 2V$_{\mathrm{O}}^{2+}$ with total charge $q=+4$.
This choice is further motivated by the single-V$_{\mathrm{O}}$ trend in Fig.~\ref{fig:fig1}(d): increasing positive charge progressively narrows the polymorph energy differences.
At this charge state the screening identifies $56$ non-PO-dominated $T_c^{*}$ configurations below the idealized $T_c$ reference, the lowest of which lies $18.4$~meV/f.u.\ below it.

\begin{figure*}[ht!]
\centering
\includegraphics[width=0.95\textwidth]{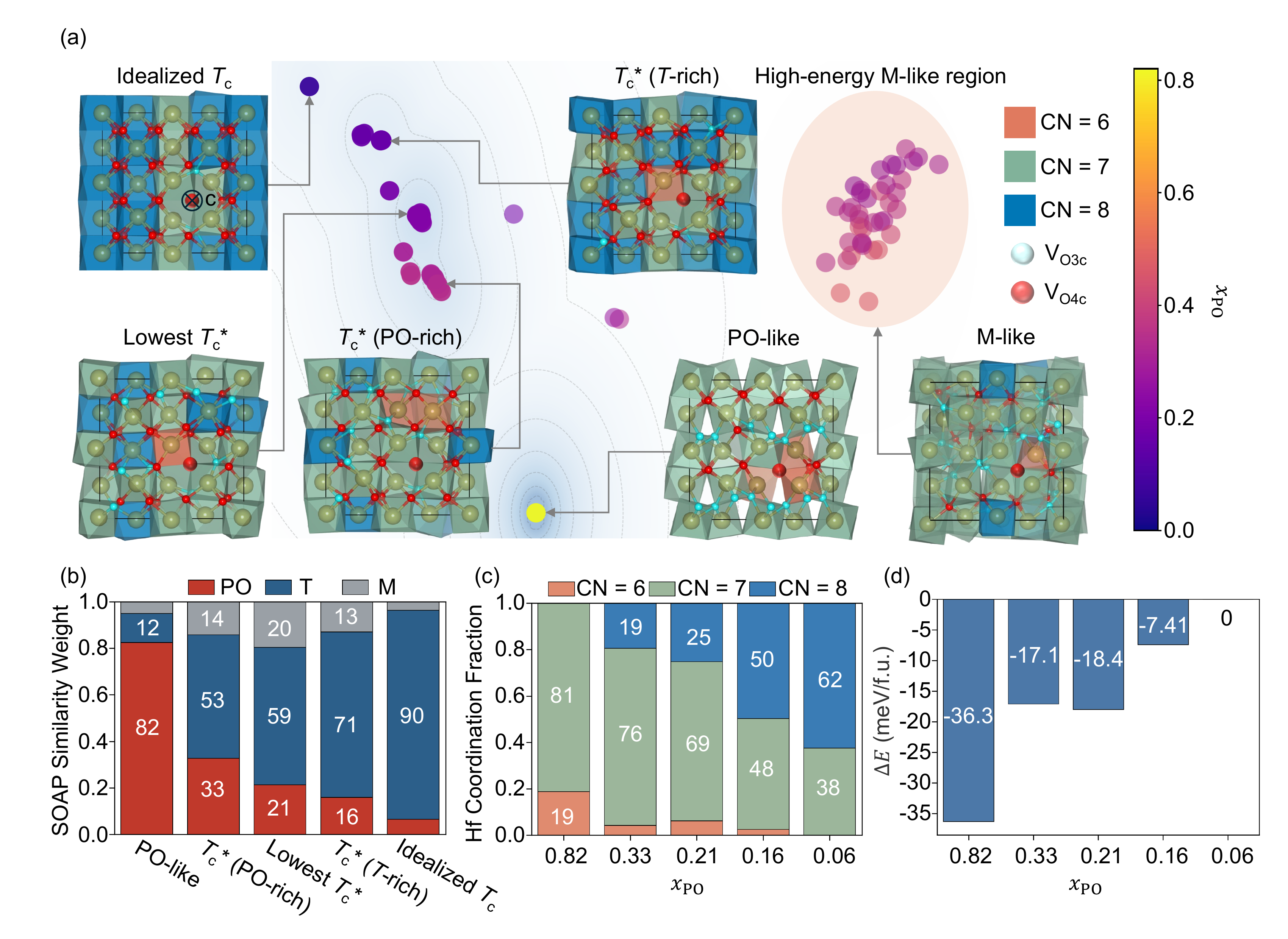}
\caption{Structural and energetic landscape of relaxed $q=+4$ $c$-axis di-V$_{\mathrm{O}}$ configurations under tetragonal-lattice confinement. (a)~SOAP-PCA map colored by the PO similarity $x_{\mathrm{PO}}=w_{\mathrm{PO}}$. Structural insets are colored by Hf coordination number, and the vacancy sites are denoted V$_{\mathrm{O3c}}$ and V$_{\mathrm{O4c}}$. (b)--(d)~SOAP similarity weights, Hf coordination fractions (CN${=}6,7,8$; $2.8$~\AA{} Hf--O cutoff), and lattice-confined relative energies of the representative configurations, referenced to the idealized $T_c$ and ordered by $x_{\mathrm{PO}}$.}
\label{fig:fig2}
\end{figure*}

To determine how these lower-energy configurations differ structurally from the idealized $T_c$ reference, we combine Hf-coordination analysis with SOAP-based structural decomposition to characterize the $T_c^{*}$ family. The Hf CN distinguishes the eightfold T-like environment from the lower-coordinated M/PO-like environments. However, CN alone cannot resolve M-like from PO-like order because M and PO both exhibit sevenfold Hf coordination and comparable Hf--O distances, while differing in the symmetry of their Hf--O networks (polar $Pca2_1$ versus centrosymmetric $P2_1/c$)~\cite{Yuan2023}. To resolve this distinction, each structure's averaged SOAP fingerprint was projected onto the simplex defined by the M, T, and PO di-V$_{\mathrm{O}}$ reference motifs, each constructed with the same $c$-axis-aligned V$_{\mathrm{O}}$--O--V$_{\mathrm{O}}$ arrangement as $T_c$ (full parameters in the Supplemental Material~\cite{SM}, Sec.~III). This projection yields barycentric weights $w_{\mathrm{M}}$, $w_{\mathrm{T}}$, and $w_{\mathrm{PO}}$, with $w_{\mathrm{M}}+w_{\mathrm{T}}+w_{\mathrm{PO}}=1$, which quantify the structural similarity to the three reference motifs. We use the PO component, $x_{\mathrm{PO}}\equiv w_{\mathrm{PO}}$, as the structural ordering coordinate.

Figure~\ref{fig:fig2}(a) places all relaxed $q=+4$ distorted configurations in a two-dimensional SOAP-PCA landscape, colored by $x_{\mathrm{PO}}$. All configurations shown, including the PO-like configuration, originate from the local-distortion search starting from the idealized $T_c$ reference. Five representative configurations are highlighted: the PO-like configuration ($x_{\mathrm{PO}}=0.82$), a PO-rich $T_c^{*}$ ($0.33$), the lowest-energy $T_c^{*}$ ($0.21$), a T-rich $T_c^{*}$ ($0.16$), and the idealized $T_c$ reference ($0.06$). $x_{\mathrm{PO}}$ near $1$ indicates PO-dominated local character, whereas the intermediate values that characterize the $T_c^{*}$ family indicate mixed character retaining contributions from M, T, and PO. The five representatives' SOAP weights, Hf coordination fractions, and lattice-confined relative energies are summarized in Figs.~\ref{fig:fig2}(b)--(d), respectively. The weights redistribute progressively from T-rich toward PO-rich local character among the selected $T_c^{*}$ representatives, with the minority M-like contribution most pronounced in the intermediate structures.
Under lattice confinement, the PO-like structure lies $36.3$~meV/f.u.\ below the idealized $T_c$ reference, lower than the other representative configurations in Fig.~\ref{fig:fig2}(d). We exclude it from the $T_c^{*}$ family, which we restrict to configurations that lie below the idealized $T_c$ reference, retain the $c$-axis V$_{\mathrm{O}}$--O--V$_{\mathrm{O}}$ alignment, and remain non-PO-dominated ($x_{\mathrm{PO}}\le0.5$). The remaining $T_c^{*}$ representatives span a continuum of mixed structural character.

The lattice-confined mixed-motif $T_c^{*}$ minimum at $x_{\mathrm{PO}}=0.21$, indicated by the arrow from the lowest $T_c^{*}$ structure in Fig.~\ref{fig:fig2}(a), has SOAP similarity weights of $w_{\mathrm{T}}=0.59$, $w_{\mathrm{PO}}=0.21$, and $w_{\mathrm{M}}=0.20$ (Fig.~\ref{fig:fig2}(b)). These redistributed SOAP weights place the selected minimum within the sampled family of mixed-motif structures, rather than near any single reference polymorph. Toward PO-like order, no sampled structures occur between the magenta PO-rich $T_c^{*}$ circle at $x_{\mathrm{PO}}=0.33$ and the yellow PO-like circle at $x_{\mathrm{PO}}=0.82$. Within the present sampling, this gap is consistent with a relatively abrupt crossover from mixed $T_c^{*}$ motifs to PO-dominated order. Figure~\ref{fig:fig2}(a) further shows that structures above the idealized $T_c$ reference cluster in a separate region and are predominantly M-like, lying at least $217$~meV/f.u.\ above it.

Independent Hf-coordination and bond-geometry analyses support the SOAP-based structural decomposition. The Hf coordination statistics in Fig.~\ref{fig:fig2}(c) follow the same structural ordering as the SOAP weights, as do the corresponding Hf--O bond-length and O--Hf--O angle distributions. From the PO-like configuration toward the idealized $T_c$ reference, CN${=}6$ and CN${=}7$ polyhedra progressively give way to CN${=}8$ coordination. For the 2V$_{\mathrm{O}}^{2+}$ complex at $q=+4$, the idealized $T_c$ reference remains CN${=}8$-dominant, whereas the lattice-confined mixed-motif $T_c^{*}$ minimum becomes CN${=}7$-dominant and exhibits a shorter and narrower Hf--O bond-length distribution, changing from $2.22\pm0.16$ to $2.19\pm0.14$~\AA{}. By contrast, the O--Hf--O angle distribution remains essentially unchanged (Supplemental Material~\cite{SM}, Fig.~S3). Shell-resolved analysis shows that sixfold- and sevenfold-coordinated Hf atoms are concentrated around the di-V$_{\mathrm{O}}$, while eightfold coordination remains more prevalent away from the defect; nevertheless, the coordination loss and Hf--O bond contraction extend into the second- and third-shell regions (Supplemental Material~\cite{SM}, Sec.~V). Consistently, the distance-resolved SOAP analysis shows that the selected $T_c^{*}$ configuration retains a T-derived background across the sampled distances, indicating that its averaged mixed-motif signal does not arise from spatially separated bulk-like regions (Supplemental Material~\cite{SM}, Sec.~III).

These local structural rearrangements are accompanied by a systematic energy gain. Within the tetragonal-lattice confinement, the representative $T_c^{*}$ configurations lie $7.4$--$18.4$~meV/f.u.\ below the idealized $T_c$ reference, with the lowest-energy mixed-motif $T_c^{*}$ configuration stabilized by $18.4$~meV/f.u.\ (Fig.~\ref{fig:fig2}(d)). To determine whether this stabilization survives the release of the lattice constraint, we subsequently relaxed both the atomic coordinates and cell parameters of the five representative configurations (Supplemental Material~\cite{SM}, Sec.~VI). Releasing the cell degrees of freedom further lowers the energy of every configuration relative to its lattice-confined counterpart and produces volume contractions of $1.8$--$3.9\%$. Importantly, after cell relaxation, all representative $T_c^{*}$ configurations remain lower in energy than the cell-relaxed idealized $T_c$ reference while retaining their characteristic structural motifs.

\begin{figure*}[t]
\centering
\includegraphics[width=0.95\textwidth]{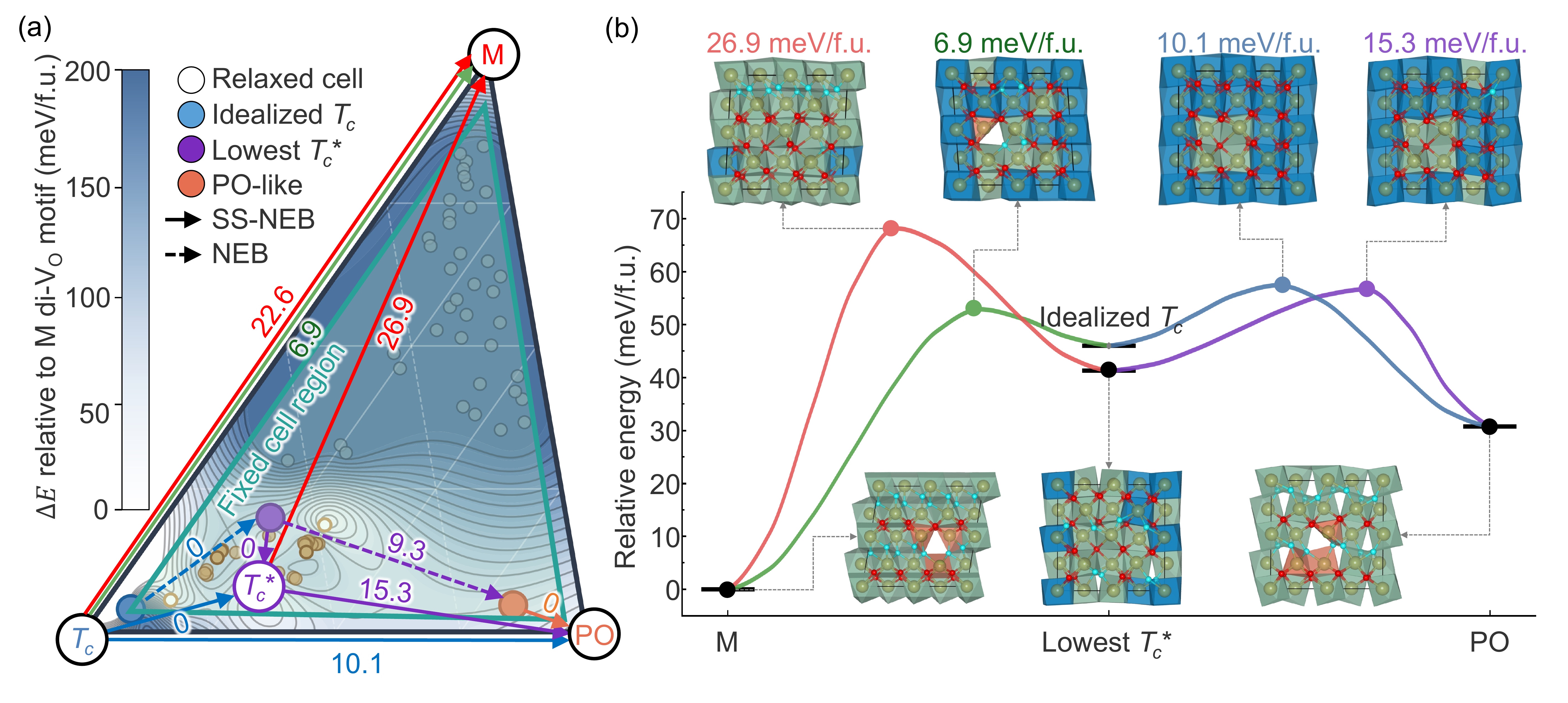}
\caption{Kinetic pathways evaluated for the fully charged $q=+4$ di-V$_{\mathrm{O}}$ motif. (a)~SOAP simplex whose circled vertices denote the idealized $T_c$, M, and PO reference motifs used for the projection; the background color and contour lines give $\Delta E$ relative to the cell-relaxed M motif. Open circles denote cell-relaxed $T_c^{*}$ configurations, whereas filled markers denote configurations under tetragonal-lattice confinement. Solid arrows show the cell-relaxed SS-NEB pathways used for the quantitative M/PO comparison, and dashed arrows show configurational connections under lattice confinement. Barriers are in meV/f.u. (b)~SS-NEB energy profiles along the competing M and PO channels, referenced to M.}
\label{fig:fig3}
\end{figure*}

To determine whether the $T_c^{*}$ motif also reshapes the polymorphic kinetics, we distinguish configurational rearrangements within the tetragonal-lattice confinement from polymorphic transformations evaluated after cell relaxation. The lattice-confined calculations preserve the lattice vectors of the tetragonal parent lattice and probe rearrangements among distinct $T_c$-derived configurations,  including fixed-cell NEB paths connecting the idealized $T_c$ reference to the lowest-energy $T_c^{*}$ configuration and this $T_c^{*}$ configuration to the PO-like configuration identified in Fig.~\ref{fig:fig2}(a). By contrast, variable-cell relaxation releases each selected di-V$_{\mathrm{O}}$ configuration from the imposed lattice strain and provides fully optimized idealized $T_c$, lowest-energy $T_c^{*}$, M, and PO structures for use as endpoints in the SS-NEB calculations.

Figure~\ref{fig:fig3}(a) projects these configurations and pathways onto the SOAP simplex whose circled vertices define the idealized $T_c$, M, and PO di-V$_{\mathrm{O}}$ reference motifs, which are free of imposed lattice strain. Open circles represent cell-relaxed $T_c^{*}$ configurations, whereas filled markers represent configurations obtained under tetragonal-lattice confinement, including the idealized $T_c$, the sampled $T_c^{*}$ family, and the PO-like motif. The background color and contour lines indicate $\Delta E$ relative to the cell-relaxed M di-V$_{\mathrm{O}}$ motif. The corresponding cell-relaxed energetics and volume changes are summarized in the Supplemental Material~\cite{SM}, Sec.~VI and Fig.~S6. As Fig.~S6 shows, di-V$_{\mathrm{O}}$ in M becomes the lowest-energy endpoint among the polymorphs once the cell is relaxed, lying $46.0$~meV/f.u.\ below the cell-relaxed idealized $T_c$, compared with $15.2$ and $4.7$~meV/f.u.\ for PO and the lowest-energy $T_c^{*}$.
This ordering reflects the bulk ground-state stability of M and is accompanied by an opposite lattice response. The $T_c$-derived structures contract on cell relaxation, whereas M expands by $1.6\%$. All configurations and pathways correspond to the same fully charged $q=+4$ complex. Dashed arrows denote configurational connections under lattice confinement, whereas solid arrows denote the cell-relaxed SS-NEB pathways used for the quantitative barrier comparison.

We first consider the kinetics of configurational rearrangements within the tetragonal-lattice confinement, shown as dashed arrows in the fixed-cell region of Fig.~\ref{fig:fig3}(a). The idealized $T_c$ reference (blue filled circle) relaxes barrierlessly into the lowest-energy $T_c^{*}$ (purple filled circle), which lies $18.4$~meV/f.u.\ lower in energy. The subsequent $T_c^{*}{\to}$PO-like NEB barrier is $9.3$~meV/f.u.\ (purple dashed arrow). The PO-like configuration (orange filled circle) is lower still in energy and relaxes barrierlessly into PO when the lattice constraint is released. By contrast, as discussed for Fig.~\ref{fig:fig2}(a), the M-like region is populated predominantly by configurations lying energetically above the idealized $T_c$ reference by at least $217$~meV/f.u., which appear as the gray circles toward the M vertex. The lattice-confined landscape is therefore energetically asymmetric, favoring relaxation toward PO-like local order while disfavoring the formation of M-like distorted configurations. This asymmetry suggests that relaxation into the distorted $T_c^{*}$ landscape preconditions the system toward PO-like local order, a possibility tested quantitatively below using cell-relaxed SS-NEB pathways.

We next examine the cell-relaxed polymorphic kinetics using variable-cell-optimized endpoints free of imposed lattice strain, shown as solid arrows in Fig.~\ref{fig:fig3}(a) and as energy profiles in Fig.~\ref{fig:fig3}(b). From the idealized $T_c$ reference (circled $T_c$ vertex), the SS-NEB calculations give a PO barrier of $10.1$~meV/f.u.\ (blue solid arrow) and two M barriers of $22.6$ and $6.9$~meV/f.u.\ (red and green solid arrows). The two M channels differ in how the oxygen rearrangement is distributed: the higher-barrier channel proceeds through a few large displacements near the divacancy, whereas the newly resolved $6.9$~meV/f.u.\ channel spreads smaller displacements across the supercell (Supplemental Material~\cite{SM}, Sec.~VII and Fig.~S8). This lower-barrier channel establishes that the idealized $T_c$ reference kinetically favors M over PO. Using the corresponding cell-relaxed $T_c^{*}$-derived structure (purple open circle) reverses this preference. As the profiles in Fig.~\ref{fig:fig3}(b) show, the $T_c^{*}{\to}$M barrier increases to $26.9$~meV/f.u.\, whereas the $T_c^{*}{\to}$PO barrier increases only to $15.3$~meV/f.u.

Taken together, Figs.~\ref{fig:fig3}(a) and~\ref{fig:fig3}(b) establish the complementary roles of the lattice-confined and cell-relaxed calculations in identifying the kinetic significance of $T_c^{*}$. The former resolves configurational rearrangements within the distorted $T_c$ landscape, revealing a lower-energy $T_c^{*}$ family and its connectivity toward the PO-like motif, whereas the latter directly compares the competing M and PO polymorphic barriers through cell-relaxed SS-NEB calculations. The $T_c{\to}T_c^{*}$ paths obtained from the fixed-cell NEB under tetragonal-lattice confinement and from the cell-relaxed SS-NEB verification both show that the selected $T_c^{*}$ configuration is kinetically accessible from the idealized $T_c$ reference without an additional activated step. The cell-relaxed SS-NEB results then reveal that this mixed-motif configuration penalizes the M branch more strongly than the PO branch, making PO the lower-barrier branch.

In conclusion, our first-principles calculations show that the idealized $T_c$ reference does not uniquely represent the $c$-axis di-V$_{\mathrm{O}}$ complex. Instead, the local-distortion search reveals a hidden structural family of lower-energy $T_c^{*}$ configurations at every charge state examined, from $q=0$ to $+4$. For the fully charged complex, the idealized reference connects barrierlessly to the $T_c^{*}$ configuration that is lowest under tetragonal-lattice confinement, lowering the energy by $18.4$~meV/f.u. This configuration retains the $c$-axis V$_{\mathrm{O}}$--O--V$_{\mathrm{O}}$ alignment while exhibiting mixed M-, T-, and PO-like character. The defect-induced structural response is not localized around the vacancy-coordinating Hf atoms, but also affects the second- and third-neighbor shells through coordination loss and Hf--O bond contraction. Representative $T_c^{*}$ configurations remain energetically below the corresponding cell-relaxed $T_c$ reference and preserve their characteristic motifs after variable-cell relaxation, demonstrating that the hidden structural family persists when the imposed lattice strain is released.

The kinetic analysis also reveals that the selected mixed-motif configuration changes the relative M and PO barriers. Within the present SS-NEB workflow, the idealized-$T_c$ paths give M and PO barriers of $6.9$ and $10.1$~meV/f.u., whereas the corresponding barriers from the selected cell-relaxed $T_c^{*}$ configuration are $26.9$ and $15.3$~meV/f.u. Although both barriers increase, the increase is substantially larger for M than for PO, making PO, rather than M, the lower-barrier branch. The branching kinetics should therefore be evaluated from the relaxed mixed-motif state, not from the idealized geometry. The combination of configurational searches and structural decomposition used here connects hidden defect configurations with kinetic branching and can be applied to other crystalline compounds with competing phases. More broadly, it offers a systematic route to investigate how defects or dopants reshape complex phase-transition kinetics by modifying local structure, metastability, and pathway selection.

\begin{acknowledgments}
This research was supported by the POSCO Science Fellowship of the POSCO TJ
Park Foundation, and the Korea Institute of Science and Technology Information
(KISTI), with supercomputing resources including technical support
(KSC-2025-CRE-0210).
\end{acknowledgments}

\bibliography{refs}

\end{document}